\documentclass[preprint2,numberedappendix]{emulateapj-rtx4}
\usepackage{graphicx,times,bm,url}
\graphicspath{{./fig/}{./png/}}
\newcommand{\EQ}{\begin{equation}}
\newcommand{\EN}{\end{equation}}
\newcommand{\EQA}{\begin{eqnarray}}
\newcommand{\ENA}{\end{eqnarray}}

\newcommand{\Fig}[1]{Figure~\ref{#1}}

\newcommand{\bra}[1]{\langle #1\rangle}

{}
{}
{}

{}
{}
{}
{}
{}
{}
{}
{}
{}
{}
{}
{}
{}
{}
{}
{}
{}

{}
{}
{}

\newcommand{\meanB}{\overline{B}}

{}

{}
{}

\newcommand{\uu}{\mbox{\boldmath $u$} {}}
\newcommand{\UU}{\mbox{\boldmath $U$} {}}

\newcommand{\bb}{\mbox{\boldmath $b$} {}}
\newcommand{\BB}{\mbox{\boldmath $B$} {}}

\newcommand{\JJ}{\mbox{\boldmath $J$} {}}

\newcommand{\AAA}{\mbox{\boldmath $A$} {}}

\newcommand{\ff}{\mbox{\boldmath $f$} {}}

\newcommand{\grav}{\mbox{\boldmath $g$} {}}
\newcommand{\nab}{\mbox{\boldmath $\nabla$} {}}
\newcommand{\SSSS}{\mbox{\boldmath ${\sf S}$} {}}

\newcommand{\DD}{{\rm D} {}}

\def\ga{\mathrel{\mathchoice {\vcenter{\offinterlineskip\halign{\hfil
$\displaystyle##$\hfil\cr>\cr\sim\cr}}}
{\vcenter{\offinterlineskip\halign{\hfil$\textstyle##$\hfil\cr>\cr\sim\cr}}}
{\vcenter{\offinterlineskip\halign{\hfil$\scriptstyle##$\hfil\cr>\cr\sim\cr}}}
{\vcenter{\offinterlineskip\halign{\hfil$\scriptscriptstyle##$\hfil\cr>\cr\sim\cr}}}}}
\def\Pm{\mbox{\rm Pr}_M}
\def\Rm{\mbox{\rm Re}_M}

\def\Rey{\mbox{\rm Re}}

\def\cs{c_{\rm s}}
\def\Pturb{P_{\rm turb}}

\def\kf{k_{\it f}} %(italics for ApJ)

\def\urms{u_{\rm rms}}

\def\etatz{\eta_{\rm t0}}

\def\tautd{\tau_{\rm td}}
\def\tauto{\tau_{\rm to}}

\def\Beq{B_{\rm eq}}
\def\Beqz{B_{\rm eq0}}

\def\half{{\textstyle{1\over2}}}

\def\onethird{{\textstyle{1\over3}}}

\newcommand{\kG}{\,{\rm kG}}

\newcommand{\yapj}[3]{ #1, {ApJ,} {#2}, #3}

\newcommand{\yan}[3]{ #1, {Astron.\ Nachr.,} {#2}, #3}

\newcommand{\yana}[3]{ #1, {A\&A,} {#2}, #3}

\newcommand{\yjetp}[3]{ #1, {Sov.\ Phys.\ JETP,} {#2}, #3}

\newcommand{\ynat}[3]{ #1, {Nature,} {#2}, #3}

\newcommand{\ysph}[3]{ #1, {Solar Phys.,} {#2}, #3}

\newcommand{\ypre}[3]{ #1, {Phys.\ Rev.\ E,} {#2}, #3}

\newcommand{\yjour}[4]{ #1, {#2}, {#3}, #4}

\newcommand{\ybook}[3]{ #1, {#2} (#3)}

\newcommand{\sapj}[1]{ #1, {ApJ}, submitted}

\newcommand{\san}[1]{ #1, {Astron. Nachr.}, submitted}
\newcommand{\smn}[1]{ #1, {MNRAS}, submitted}

\begin{document}

\title{Detection of negative effective magnetic pressure instability in turbulence simulations}

\author{Axel Brandenburg$^{1,2}$, Koen Kemel$^{1,2}$,
Nathan Kleeorin$^{3,1}$, Dhrubaditya Mitra$^1$, and Igor Rogachevskii$^{3,1}$
}
\affil{
$^1$NORDITA, AlbaNova University Center, Roslagstullsbacken 23,
SE-10691 Stockholm, Sweden \\
$^2$Department of Astronomy, AlbaNova University Center,
Stockholm University, SE-10691 Stockholm, Sweden\\
$^3$Department of Mechanical
Engineering, Ben-Gurion University of the Negev, POB 653,
Beer-Sheva 84105, Israel
}

\date{Received 2011 September 6; accepted 2011 September 19; published 2011 October 3}
\submitted{}
\journalinfo{The Astrophysical Journal Letters, 740:L50 (4pp), 2011 October 20\hfill doi:10.1088/2041-8205/740/2/L50}

\begin{abstract}
We present the first demonstration of the negative effective
magnetic pressure instability in direct numerical simulations of
stably stratified, externally forced, isothermal hydromagnetic turbulence in the regime of large plasma beta.
By the action of this instability, an initially uniform horizontal
magnetic field forms flux concentrations whose scale is large
compared to the turbulent scale.
We further show that the magnetic energy of these large-scale structures
is only weakly dependent on the magnetic Reynolds number,
provided its value is large enough for the instability to be excited.
Our results support earlier mean-field calculations and analytic work
which identified this instability.
Applications to the formation of active regions in the Sun are discussed.
\end{abstract}

\section{Introduction}

The solar convection zone is highly turbulent and
mixing is expected to be efficient.
Nevertheless, the Sun displays coherent structures encompassing
many turbulent eddy scales.
A well-known example is the large-scale magnetic field of the Sun that is
antisymmetric about the equator and shows a 22 year solar cycle \citep{SV86}.
Another prominent example in the Sun is the emergence of active regions.
It is generally believed that active regions are
the result of some non-axisymmetric instability
of $\sim100\kG$ magnetic fields in the tachocline \citep{GD00,CDG03,PM07}.
However, the existence of such strong fields remains debatable \citep{B05}.

A powerful tool for understanding the emergence of such large-scale
structures from a turbulent background is mean-field dynamo theory
\citep{Mof78,Par79,KR80}.
With the advent of powerful computers and numerical simulation tools,
it has become possible to confront many of the mean-field predictions
with direct simulations \citep{BS05}.
Here we consider the idea that statistically steady, stratified, hydromagnetic
turbulence with an initially uniform magnetic field is unstable to the
negative effective magnetic pressure instability (NEMPI).
This instability is caused by the suppression of turbulent
hydromagnetic pressure (the isotropic part of combined Reynolds and Maxwell stresses) by the mean magnetic field \citep{KRR90,RK07}.
At large Reynolds numbers and for sub-equipartition magnetic fields,
the negative
turbulent contribution can become so large that the
effective mean magnetic pressure
(the sum of turbulent and non-turbulent contributions)
appears negative.
In a stratified medium,
this results in the excitation of NEMPI that causes formation
of large-scale inhomogeneous magnetic structures.
NEMPI is similar to the large-scale dynamo instability, except that it
only redistributes the total magnetic flux, creating large-scale
concentrated magnetic flux regions at the expense of turbulent
kinetic energy.

Historically, the magnetic suppression of the Reynolds stress
was first found by \cite{Rae74} and \cite{Rue74}.
Later, \cite{Rue86} considered the Maxwell stress and found
the mean effective magnetic tension to be suppressed by mean fields.
However, these calculations were based on quasi-linear theory
which is only valid at low fluid and magnetic Reynolds numbers.
\cite{KRR90,KMR96} considered the combined Reynolds and Maxwell
stresses at large Reynolds numbers and found
a sign reversal of the effective mean magnetic pressure.
This result is based on the $\tau$ approximation,
and has been corroborated using
the renormalization procedure \citep{KR94}.

\begin{figure*}\begin{center}
\includegraphics[width=\textwidth]{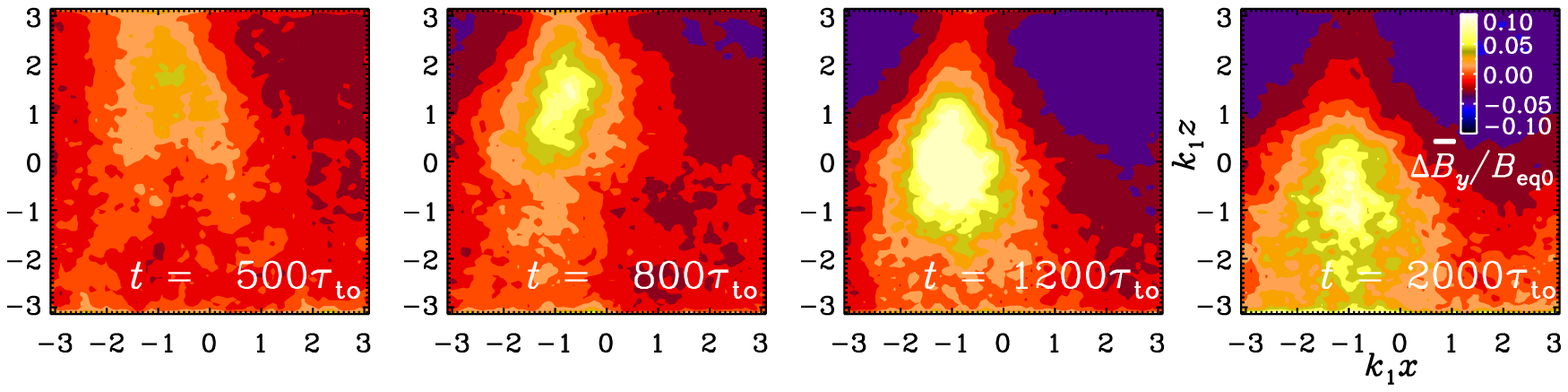}
%\plotone{f1}
\end{center}\caption[]{
$\Delta\meanB_y/\Beqz$ in the $xz$ plane for $\Rm=6$ and $B_0/\Beqz=0.05$,
showing a descending ``potato sack'' structure.
Time is in units of $\tauto$ (lower right).
}\label{pbymxztn}\end{figure*}

The magnetic suppression of the combined Reynolds and Maxwell stresses
is quantified in terms of new turbulent mean-field coefficients
that relate the components of the sum of Reynolds and Maxwell stresses
to the mean magnetic field.
These coefficients depend on the magnetic field and have now been determined
in direct numerical simulations (DNS)
for a broad range of different cases, including unstratified forced turbulence
\citep{BKR10}, isothermally stratified forced turbulence
\citep[][hereafter BKKR]{BKKR11},
and turbulent convection \citep{KBKMR11}.
These simulations have clearly demonstrated that the mean effective
magnetic pressure is negative for magnetic field strengths below
about half the equipartition field strength.
However, these DNS studies did not find the actual instability.

With a quantitative parameterization in place, it became possible
to build mean-field models of stratified turbulence which clearly
demonstrate exponential growth and saturation of NEMPI.
In view of applications to the formation of active regions in the Sun,
such simulations were originally done for an adiabatically stratified
layer \citep{BKR10}.
In addition, mean-field studies showed the existence of NEMPI even
for isothermal stably stratified layers \citep{KBKR11}.
This last result turned out to be important because it paved
the way for this Letter where we demonstrate NEMPI through DNS.
Once we establish the physical reality of this effect, it would be
important to apply it to realistic solar models which include proper
boundary conditions, realistic stratification, convective flux,
and radiation transport.
However, at this stage it is essential to isolate NEMPI as a physical
effect under conditions that are as simple as possible.

\section{The model}
\label{DNSmodel}

We consider a domain of size $L_x\times L_y\times L_z$ in
Cartesian coordinates, $(x,y,z)$, with periodic boundary conditions in
the $x$ and $y$ directions and stress-free perfectly conducting
boundaries at top and bottom ($z=\pm L_z/2$).
The volume-averaged density is therefore constant in time and equal to its initial value, $\rho_0=\bra{\rho}$.
We solve the equations for the velocity $\UU$,
the magnetic vector potential $\AAA$, and the density $\rho$,
\begin{equation}
\rho{\DD\UU\over\DD t}=\JJ\times\BB-\cs^2\nab\rho+\nab\cdot(2\nu\rho\SSSS)
+\rho(\ff+\grav),
\end{equation}
\begin{equation}
{\partial\AAA\over\partial t}=\UU\times\BB+\eta\nabla^2\AAA,
\end{equation}
\begin{equation}
{\partial\rho\over\partial t}=-\nab\cdot\rho\UU,
\end{equation}
where $\nu$ is kinematic viscosity, $\eta$ is magnetic diffusivity,
$\BB=\BB_0+\nab\times\AAA$ is the magnetic field,
$\BB_0=(0,B_0,0)$ is the imposed uniform field,
$\JJ=\nab\times\BB/\mu_0$ is the current density,
$\mu_0$ is the vacuum permeability,
${\sf S}_{ij}=\half(U_{i,j}+U_{j,i})-\onethird\delta_{ij}\nab\cdot\UU$
is the traceless rate of strain tensor, and commas denote
partial differentiation.
The forcing function $\ff$ consists of
random, white-in-time, plane non-polarized waves with an average
wavenumber $\kf=15\,k_1$,
where $k_1=2\pi/L_z$ is the lowest wavenumber in the domain.
The forcing strength is such that the turbulent rms velocity is approximately
independent of $z$ with $\urms=\bra{\uu^2}^{1/2}\approx0.1\,\cs$.
The gravitational acceleration $\grav=(0,0,-g)$ is chosen such that
$k_1 H_\rho=1$, which leads to a density contrast between bottom and top
of $\exp(2\pi)\approx535$.
Here, $H_\rho=\cs^2/g$ is the density scale height.

Our simulations are characterized by the
fluid Reynolds number $\Rey\equiv\urms/\nu\kf$,
the magnetic Prandtl number $\Pm=\nu/\eta $ and
the magnetic Reynolds number $\Rm\equiv \Rey \, \Pm$.
Following earlier work \citep{BKKR11}, we choose
$\Pm = 0.5$ and $\Rm$ in the range $0.7$--$74$.
The magnetic field is expressed in units of the local
equipartition field strength near the top, $\Beq=
\sqrt{\mu_0\rho} \, \urms$, while
$B_0$ is specified in units of the
averaged value, $\Beqz=\sqrt{\mu_0\rho_0} \, \urms$.
We monitor $\Delta\meanB_y=\meanB_y-B_0$, where $\meanB_y$ is an
average over $y$ and a certain time interval $\Delta t$.
Time is expressed in eddy turnover times, $\tauto=(\urms\kf)^{-1}$.
Occasionally, we also consider the turbulent-diffusive timescale,
$\tautd=(\etatz k_1^2)^{-1}$, where $\etatz=\urms/3\kf$ is
the estimated turbulent magnetic diffusivity.
Another diagnostic quantity is the rms magnetic field in the $k=k_1$
Fourier mode, $B_1$, which is here taken as an average over
$2\leq k_1z\leq3$, and is close to the top at $k_1 z=\pi$.
(Note that $B_1$ does not include the imposed field $B_0$ at $k=0$.)

\begin{figure}\begin{center}
\includegraphics[width=\columnwidth]{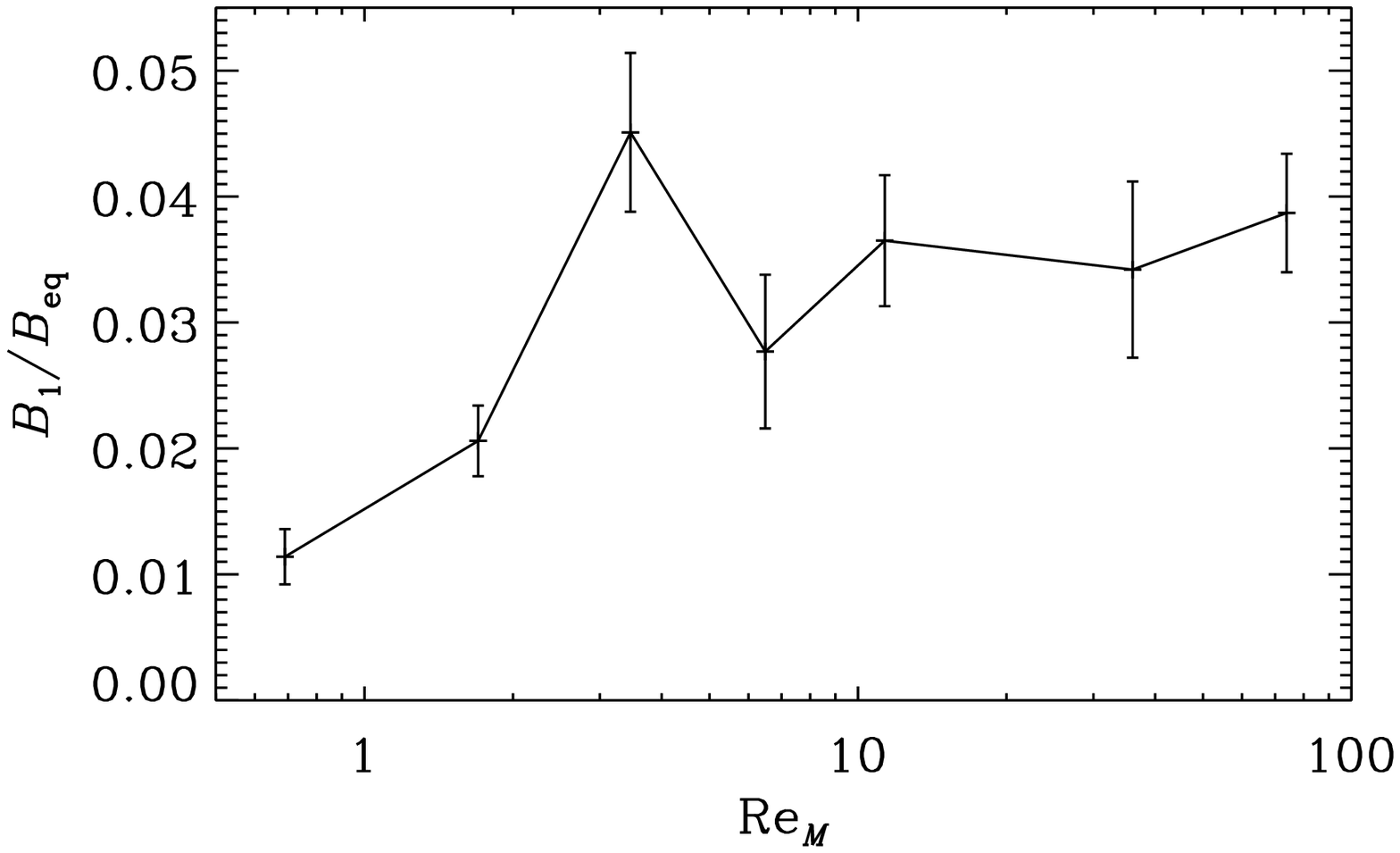}
%\plotone{f2}
\end{center}\caption[]{
Dependence of $B_1/\Beq$ (averaged over $2\leq k_1z\leq3$)
on $\Rm$ for $B_0/\Beqz=0.05$.
}\label{psummaryR}\end{figure}

\begin{figure*}\begin{center}
\includegraphics[width=\textwidth]{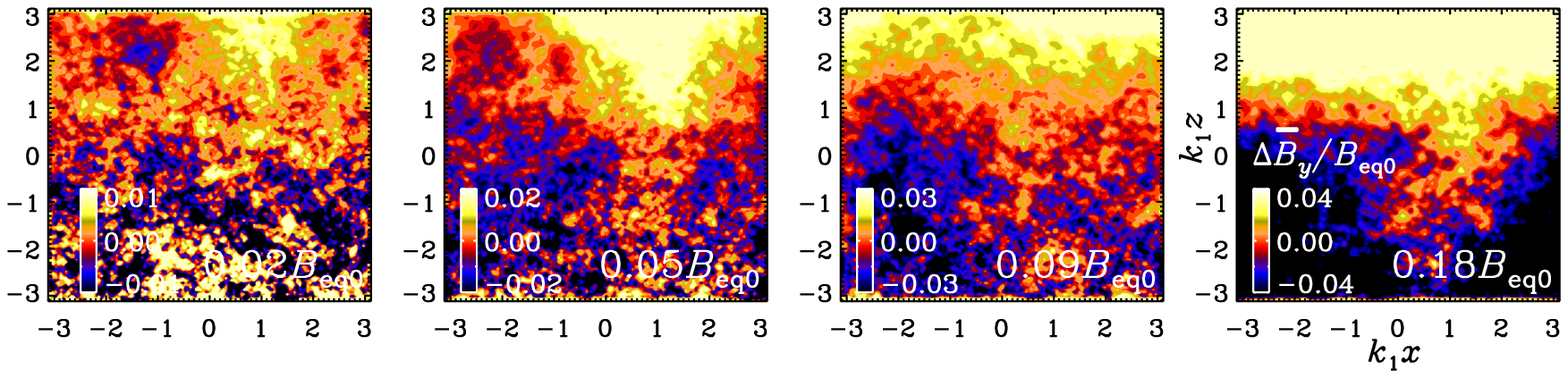}
%\plotone{f3}
\end{center}\caption[]{
$\Delta\meanB_y/\Beqz$ in the $xz$ plane for $\Rm=36$ and
$B_0/\Beqz=0.02$, 0.05, 0.09, and 0.18.
}\label{pcomp_slice}\end{figure*}

The simulations are performed with the {\sc Pencil Code},%
\footnote{{\tt http://pencil-code.googlecode.com}}
which uses sixth-order explicit finite differences in space and a
third-order accurate time stepping method.
We use numerical resolutions of $128^3$ and $256^3$ mesh points
when $L_x=L_y=L_z$, and $1024\times128^2$ when $L_x=8L_y=8L_z$.
To capture mean-field effects on the slower turbulent-diffusive
timescale, which is $\tautd/\tauto=3\kf^2/k_1^2$ times slower than
the dynamical timescale, we perform simulations for several thousand
turnover times.

\section{Results}

The NEMPI phenomenon is already quite pronounced at intermediate values
of $\Rm\ga3$; see \Fig{pbymxztn},
where we show $B_y$ averaged over $y$ and $\Delta t\approx80\tauto$
(denoted by $\meanB_y$)
at selected times during the first 2000 turnover times.
The $\Rm$ dependence of $B_1/\Beq$ is shown in \Fig{psummaryR}.
For $\Rm>3$, the
$\Rm$ dependence is relatively weak, which suggests that NEMPI is
a genuine high Reynolds number effect.

The results for the case of $\Rm=6$ show strong similarities to earlier
mean-field simulations.
During the first 500 turnover times, flux concentrations form first
near the surface, but at later times the location of the peak magnetic field moves
gradually downward.
This phenomenon is a direct consequence of the negative effective magnetic
pressure, making such structures heavier than their surroundings.
Their shape resembles that of a falling ``potato sack'' and
has been seen in numerous mean-field calculations
during the nonlinear stage of NEMPI \citep{BKR10,KBKMR11}.
Using the technique described in BKKR, we have found that
for $\Rm\ga1.1$ and $\Pm=0.5$,
the effective magnetic pressure has a negative minimum.

As expected from theory and mean-field calculations,
NEMPI is only excited in a certain range of field strengths.
In particular, only for $B_0/\Beqz$ between 0.02 and 0.2
do we see large-scale magnetic structures.
This is shown in \Fig{pcomp_slice}, where we see $B_y$,
again averaged over $y$ and a time interval $\Delta t\approx800\tauto$,
in which the field is statistically steady.
The clearest flux structure formations are seen for $B_0/\Beqz\approx0.05$.
However, even for this case the flux concentrations are barely
visible in a single snapshot.
This has been one of the reasons why NEMPI has not
been noticed before in DNS.
An additional handicap was that the simulations of BKKR
used a smaller scale separation ratio of only 5, which is
nevertheless still sufficient
for determining the governing mean-field coefficients and
allows one to reach larger values of $\Rm$.

\begin{figure}\begin{center}
\includegraphics[width=\columnwidth]{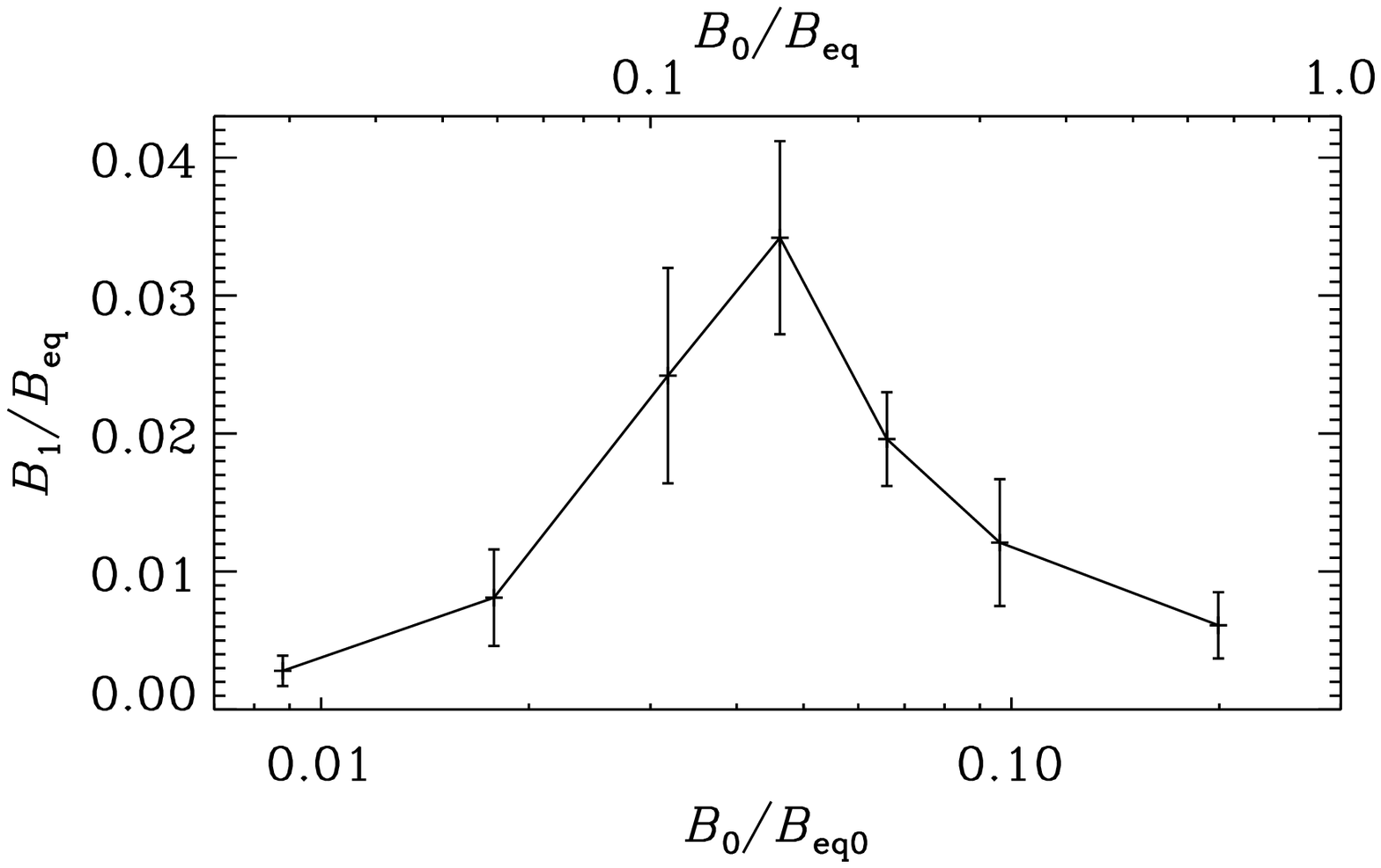}
%\plotone{f4}
\end{center}\caption[]{
Bifurcation diagram showing $B_1/\Beq$ vs.\ $B_0/\Beqz$
(and vs.\ $B_0/\Beq$ on the upper abscissa) for $\Rm=36$.
}\label{psummary}\end{figure}

\begin{figure}\begin{center}
\includegraphics[width=\columnwidth]{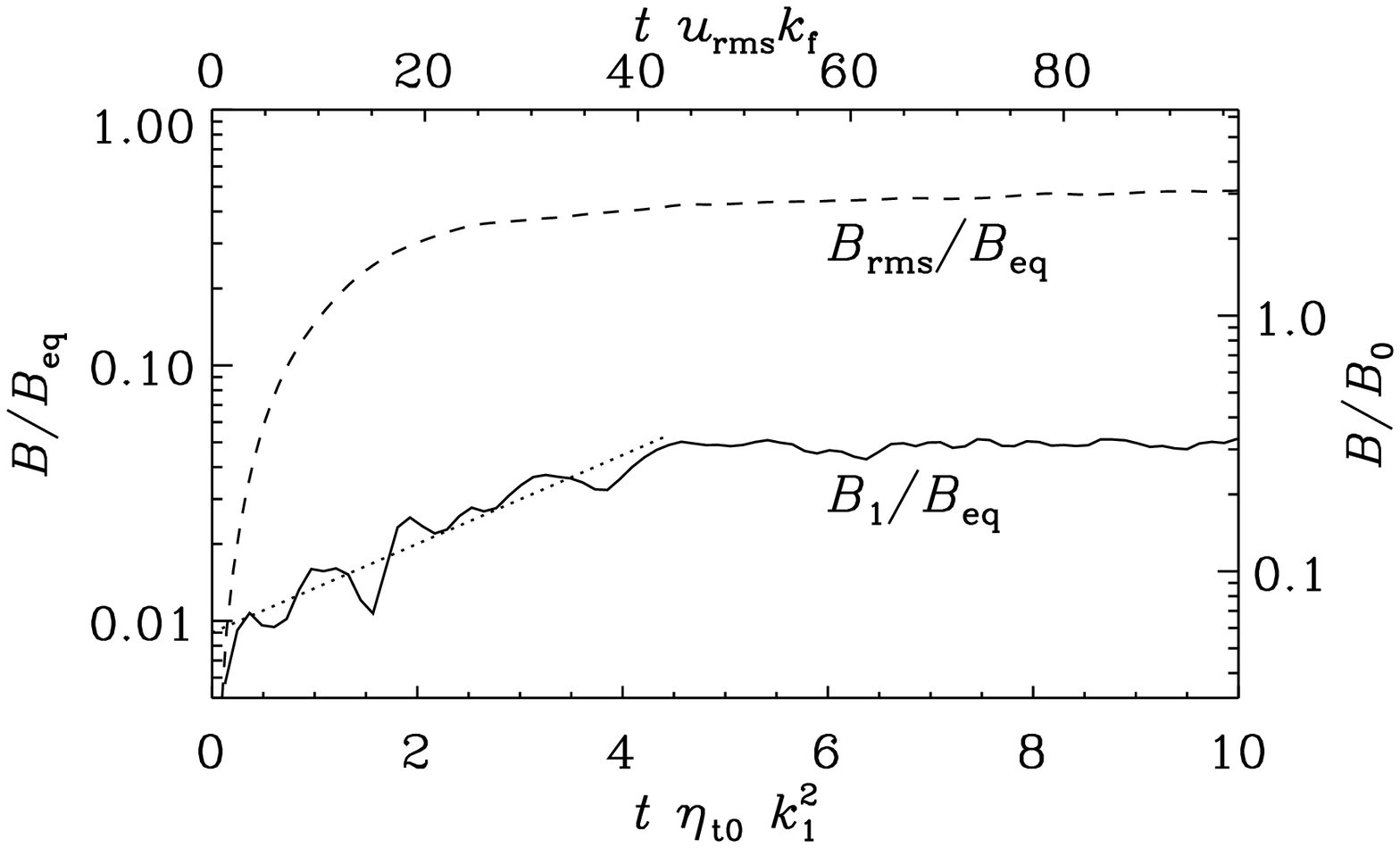}
%\plotone{f5}
\end{center}\caption[]{
Slow exponential growth of the mean magnetic field in $2\leq k_1z\leq3$
for $B_0/\Beqz=0.05$, corresponding to $B_0/\Beq=0.16$, for $\Rm=36$
(solid line).
The growth rate is $\approx0.4\etatz k_1^2$ (dotted line).
Note that the total rms field, $B_{\rm rms}/\Beq$ (dotted line)
saturates much faster, as indicated by to upper abscissa.
}\label{ppbymxztm}\end{figure}

In \Fig{psummary} we plot $B_1/\Beq$ as a function of $B_0/\Beqz$,
showing a peak at $B_0/\Beqz\approx0.05$.
We recall that $\Beq$ applies here to the location
$2\leq k_1z\leq3$ where $B_1$ has been evaluated, and there we have
$\Beq/\Beqz\approx0.3$.
The fact that large-scale flux concentrations develop only for
a certain range of imposed field strengths supports our interpretation
that they are caused by NEMPI and not, for example,
by some yet unknown dynamo mechanism.
In all these cases, $\BB-\BB_0$ grows rapidly and reaches a
saturation field strength that is independent of $\BB_0$, provided $\Rm\ge35$.
This suggests that this field is produced by small-scale dynamo action
and not just by field line tangling.
Another piece of evidence of the physical reality of NEMPI is
shown in \Fig{ppbymxztm}, where we see that $B_1$ does indeed increase
exponentially for the first 2000 turnover times, corresponding to about
$3\tautd$.
The growth rate is $\approx0.4\etatz k_1^2$, which is much less than
$\tauto^{-1}$, but entirely compatible with mean-field calculations
\citep[BKKR;][]{KBKMR11}.

Finally, to investigate the effects of the domain aspect ratio
on the instability, we perform a calculation with $B_0/\Beqz=0.05$,
$\Rm=36$, and change $L_x$ to $16\pi/k_1$.
We find that the most unstable
mode has a wavelength approximately equal to $L_z \approx6H_\rho$.
This result is also in agreement with mean-field
models \citep[e.g., Fig.~14 of][]{KBKMR11}.
The large-scale flux concentrations have an amplitude of only
$\approx0.1\Beq$ and are therefore not seen in single snapshots,
where the field reaches peak strengths comparable to $\Beq$.
Furthermore, as for any linear instability, the flux concentrations
form a repetitive pattern, but this might be an artifact of
idealized conditions.

\section{Conclusions and discussion}

The present simulations have, for the first time, demonstrated conclusively
that NEMPI can operate
in hydromagnetic turbulence under proper conditions, namely,
strong stratification, sufficient scale separation (here $\kf/k_1=15$), and a
mean field in an optimal range (here $\approx0.15\Beq$; see \Fig{psummary}).
This instability has so far only been seen in mean-field simulations.
By contrast, the present simulations are completely free of
any mean-field consideration.

The instability is argued to be a consequence
of the reduction of turbulent hydromagnetic pressure
by a mean magnetic field and can be understood as follows \citep{KMR96}.
The combined Reynolds and Maxwell stress is $\overline{\rho u_i u_j}
-\overline{b_i b_j}/\mu_0+\delta_{ij}\overline{\bb^2}/2\mu_0$,
where $\uu$ and $\bb$ are velocity and magnetic fluctuations,
respectively, and overbars indicate averaging.
For isotropic turbulence, the turbulent hydromagnetic pressure is then
$\Pturb={\textstyle{1\over3}}(\overline{\rho \uu^2}
+\overline{\bb^2}/2\mu_0)$.
On the other hand, the total turbulent energy
$E_{\rm turb} \equiv \half(\overline{\rho\uu^2}+\overline{\bb^2}/\mu_0)$,
is nearly conserved
because a uniform mean magnetic field does not perform any work;
see \cite{BKR10} for a numerical demonstration.
The presence of an additional 1/2 factor in front of the
$\overline{\bb^2}/\mu_0$ term in the expression for $\Pturb$,
but not in that for $E_{\rm turb}$, implies that the
generation of magnetic fluctuations results in a reduction of $\Pturb$,
i.e.\ $\Pturb={\textstyle{1\over3}}(2E_{\rm turb}-\overline{\bb^2}/2\mu_0)$.
For anisotropic turbulence this negative contribution becomes larger.
This physical effect is independent of stratification,
but to obtain an instability one needs strong stratification.

We speculate that, in the solar context,
NEMPI plays a role in formation of active regions
from mean fields generated by the solar dynamo.
Let us now ask whether this instability alone can describe the
formation of active regions at the solar surface.
Clearly, the flux concentrations we observe are not strong enough
to be noticeable without averaging, while
the active regions in the Sun are seen without
averaging.
This suggests that there may be additional mechanisms at work.
One possibility is that of the magnetic suppression of the
convective heat flux that has been invoked to explain the formation
of sunspots \citep{KM00}.

When the mean magnetic field becomes larger than the equipartition field strength, i.e., when NEMPI does not work, and the characteristic spatial scale
of the magnetic field is smaller than the density height,
it is instead the Parker magnetic buoyancy instability
\citep{Par66} that is excited.

The presence of a vertical field might also have a strong effect.
Indeed simulations of convection with an imposed vertical field
have produced a segregation into magnetized and unmagnetized
regions \citep{Tao98,KKWM10} with formation of flux concentrations
strong enough to be noticeable even without averaging.
On the other hand NEMPI might become more powerful at stronger stratification.
Increased stratification clearly has an enhancing effect on the growth
rate \citep{KBKR11}, but the effect on the saturation level has not yet
been quantified.
Furthermore, the interplay between NEMPI and the other
effects also needs to be investigated.

Our work has established a close link between
what can be expected from mean-field studies and what actually
happens in DNS.
This correspondence is particularly important because DNS
cannot reach solar parameters in any conceivable future.
Hence a deeper understanding of solar convection can
only emerge
by studying mean-field models on the one hand and to determine turbulent
mean-field coefficients from simulations on the other hand.
This concerns not only the dependence of the mean-field coefficients
on parameters
such as magnetic Reynolds and Prandtl numbers and scale separation
ratio, but also the details of the source of turbulence.
In particular, it has already been shown that the negative effective
magnetic pressure effect is not unique to forced turbulence,
but it also occurs in turbulent convection \cite{KBKMR11},
and thus in an unstably stratified layer.
We emphasize that the present work  demonstrates
the predictive power of mean-field theory at an advanced level
where quasi-linear theory fails \citep{RKS11}
but the spectral $\tau$ approach \citep{RK07} has proven useful.

More work using mean-field models is needed to elucidate details of
the mechanism of NEMPI.
For example, naive thinking suggests that the onset of NEMPI should
occur at the depth where the effective magnetic pressure is minimum,
but both mean-field models and DNS show that this is not the case.
At least at early times, NEMPI appears
most pronounced at the top of the domain, while the
effective magnetic pressure is usually most negative at the bottom.
On the other hand, the instability is a global one and local
considerations such as these are not always meaningful.
Another question is what happens when the imposed field is replaced
by a dynamo-generated one.
In that case, the turbulence may be helical and new terms involving
current density can occur in the expression for the mean-field stress.
Again, such possibilities are best studied using first the mean-field approach.

\acknowledgments

We acknowledge the NORDITA dynamo programs of 2009 and 2011 for
providing a stimulating scientific atmosphere.
Computing resources provided by the Swedish National Allocations Committee
at the Center for Parallel Computers at the Royal Institute of Technology in
Stockholm and the High Performance Computing Center North in Ume{\aa}.
This work was supported in part by the European Research Council
under the AstroDyn Research Project No.\ 227952.

\end{document}